\documentclass[prl,twocolumn,aps,floatfix]{revtex4}

\usepackage{graphicx}
\newcommand{\ba}{\begin{eqnarray}}
\newcommand{\ea}{\end{eqnarray}}

\begin{document}
\title{Evidence for tetrahedral symmetry in $^{16}$O}
\author{R. Bijker$^1$ and F. Iachello$^{2}$}
\address{$^1$Instituto de Ciencias Nucleares, 
Universidad Nacional Aut\'onoma de M\'exico, \\ 
Apartado Postal 70-543, 04510 M\'exico, D.F., M\'exico \\
$^2$ Center for Theoretical Physics, Sloane Laboratory, \\
Yale University, New Haven, CT 06520-8120, U.S.A.}

\begin{abstract}
We derive the rotation-vibration spectrum of a 4$\alpha$ configuration with 
tetrahedral symmetry, ${\cal T}_d$, and show evidence for the occurrence of this 
symmetry in the low-lying spectrum of $^{16}$O. All vibrational states with $A$, 
$E$ and $F$ symmetry appear to have been observed, as well as the rotational 
bands with $L^P=0^+$, $3^-$, $4^+$, $6^+$ on the $A$ states, and part of the 
rotational bands built on the $E$, $F$ states. We derive analytic expressions 
for the form factors and $B(EL)$ values of the ground state rotational band and 
show that the measured values support the tetrahedral symmetry of this band. 
\end{abstract}


\maketitle

The cluster structure of light nuclei is a long standing problem which goes back to 
the early days of nuclear physics \cite{Wheeler}. Recent experimental developments have shown 
that the low-lying states of $^{12}$C can be described as rotation-vibration of a $3\alpha$ 
cluster with ${\cal D}_{3h}$ symmetry (equilateral triangle) \cite{Itoh,Freer,Gai,Marin}. 
Departures from a rigid cluster structure appear to be moderate in size and can be accounted 
for by perturbation theory. In this article, we show that the low-lying states of $^{16}$O 
can be described as rotation-vibration of a $4\alpha$ cluster with ${\cal T}_d$ symmetry 
(tetrahedral). The suggestion that $^{16}$O has a tetrahedral $4\alpha$ structure goes 
back many years \cite{Dennison,Kameny,Brink1,Brink2,Robson}. 
However, clear signatures could not be identified. We take advantage of the algebraic cluster 
model (ACM) \cite{BI1,BI2} to produce the rotation-vibration spectrum of an object with ${\cal T}_d$ 
symmetry and compare with the observed spectrum. We also derive an analytic expression for the 
$B(EL)$ values along the ground state rotational band. A comparison with the experimental values 
of the energy spectrum and electromagnetic transitions provides strong evidence for tetrahedral 
symmetry in $^{16}$O. 

The algebraic cluster model is a description of cluster states as representations of a 
$U(\nu+1)$ group where $\nu$ is the number of space degrees of freedom \cite{BI1,BI2}. 
In Ref.~\cite{BI1,BI2}, we described three-body clusters, where the number of degrees of 
freedom (after removal of the center of mass) is $\nu=3n-3=6$, in terms of the algebra of 
$U(7)$. The space degrees of freedom are there the Jacobi coordinates  
$\vec{\rho} = \left( \vec{r}_{1} - \vec{r}_{2} \right) /\sqrt{2}$ and 
$\vec{\lambda} = \left( \vec{r}_{1} + \vec{r}_{2} - 2\vec{r}_{3} \right)/\sqrt{6}$, 
where $\vec{r}_i$ are the coordinates of the three $\alpha$ particles ($i=1,2,3$). 
We describe four-body clusters with $\nu=3n-3=9$ in terms of the algebra of $U(10)$. 
The space degrees of freedom are here three Jacobi vectors, 
$\vec{\rho} = \left( \vec{r}_{1} - \vec{r}_{2} \right) /\sqrt{2}$, 
$\vec{\lambda} = \left( \vec{r}_{1} + \vec{r}_{2} - 2\vec{r}_{3} \right)/\sqrt{6}$ and 
$\vec{\eta} = \left( \vec{r}_{1} + \vec{r}_{2} + \vec{r}_{3} - 3\vec{r}_{4} \right)/\sqrt{12}$,  
where $\vec{r}_i$ are the coordinates of the four $\alpha$ particles ($i=1,\ldots,4$). 
The algebra of $U(10)$ is constructed by introducing three vector bosons, $b_{\rho}$, 
$b_{\lambda}$ and $b_{\eta}$, together with an auxiliary scalar boson, $s$. 
The bilinear products of creation and annihilation operators generate the algebra $U(10)$
\ba
b_{\rho ,m}^{\dagger }, \; b_{\lambda ,m}^{\dagger }, \; b_{\eta,m}^{\dagger}, \; s^{\dagger} 
\;\equiv\; c_{\alpha}^{\dagger } \hspace{1cm} (m=0,\pm 1) ~, 
\nonumber\\
{\cal G} \;:\; G_{\alpha \beta } \;=\; c_{\alpha}^{\dagger} c_{\beta } 
\hspace{1cm} (\alpha ,\beta =1, \ldots,10) ~. 
\nonumber
\ea
The creation and annihilation operators for vector bosons 
($b_{\rho,m}^{\dagger}$, $b_{\lambda,m}^{\dagger}$, $b_{\eta,m}^{\dagger }$ and 
$b_{\rho,m}$, $b_{\lambda,m}$, $b_{\eta,m}$) represent the second quantized form 
of the Jacobi coordinates and their canonically conjugate momenta, 
while the auxiliary scalar boson is introduced in order to
construct the spectrum generating algebra. 
The energy levels can be obtained by diagonalizing the Hamiltonian $H$. In this article, 
we consider clusters composed of four identical particles (4$\alpha$), for which $H$ must 
be invariant under the permutation group $S_4$. The most general one- and two-body 
Hamiltonian that describes the relative motion of four identical particles, is a scalar 
under $S_4$, is rotationally invariant, and conserves parity as well as the total number of 
bosons is given by \cite{RB1,RB2} 
\ba
H &=& \epsilon_{0} \, s^{\dagger} \tilde{s}
- \epsilon_{1} \, (b_{\rho}^{\dagger} \cdot \tilde{b}_{\rho} 
+ b_{\lambda}^{\dagger} \cdot \tilde{b}_{\lambda}  
+ b_{\eta}^{\dagger} \cdot \tilde{b}_{\eta})
\nonumber\\ 
&& + u_0 \, s^{\dagger} s^{\dagger} \tilde{s} \tilde{s}  
- u_1 \, s^{\dagger} ( b_{\rho}^{\dagger} \cdot \tilde{b}_{\rho} 
+ b_{\lambda}^{\dagger} \cdot \tilde{b}_{\lambda}  
+ b_{\eta}^{\dagger} \cdot \tilde{b}_{\eta} ) \tilde{s} 
\nonumber\\
&& + v_0 \, \left[ ( b_{\rho}^{\dagger} \cdot b_{\rho}^{\dagger} 
+ b_{\lambda}^{\dagger} \cdot b_{\lambda}^{\dagger}  
+ b_{\eta}^{\dagger} \cdot b_{\eta}^{\dagger} ) \tilde{s} \tilde{s} + {\rm h.c.} \right]
\nonumber\\
&& + \sum_{L=0,2} a_{L} \, \left[ [ 2 b_{\rho}^{\dagger} b_{\eta}^{\dagger}
+ 2\sqrt{2} \, b_{\rho}^{\dagger} b_{\lambda}^{\dagger} ]^{(L)} 
\cdot [ {\rm h.c.} ]^{(L)} \right.
\nonumber\\
&& \hspace{1cm} + [ 2 b_{\lambda}^{\dagger} b_{\eta}^{\dagger} 
+ \sqrt{2} \, ( b_{\rho}^{\dagger} b^{\dagger}_{\rho}  
- b_{\lambda}^{\dagger} b_{\lambda}^{\dagger} ) ]^{(L)} \cdot [ {\rm h.c.} ]^{(L)} 
\nonumber\\
&& \hspace{1cm} \left. + [ b_{\rho}^{\dagger} b_{\rho}^{\dagger} 
     + b_{\lambda}^{\dagger} b_{\lambda}^{\dagger} 
   - 2 b_{\eta}^{\dagger} b_{\eta}^{\dagger} ]^{(L)} \cdot [ {\rm h.c.} ]^{(L)} \right]
\nonumber\\
&& + \sum_{L=0,2} c_{L} \, \left[
[ -2\sqrt{2} \, b_{\rho}^{\dagger} b_{\eta}^{\dagger}
+ 2 b_{\rho}^{\dagger} b_{\lambda}^{\dagger} ]^{(L)} \cdot [ {\rm h.c.} ]^{(L)} \right.
\nonumber\\
&& \hspace{1cm} + \left. [ -2\sqrt{2} \, b_{\lambda}^{\dagger} b_{\eta}^{\dagger} 
+ ( b_{\rho}^{\dagger} b_{\rho}^{\dagger} 
- b_{\lambda}^{\dagger} b_{\lambda}^{\dagger} ) ]^{(L)} \cdot [ {\rm h.c.} ]^{(L)} \right]
\nonumber\\
&& + c_1 \, \left[ 
  ( b_{\rho}^{\dagger} b_{\lambda}^{\dagger} )^{(1)} \cdot 
  ( \tilde{b}_{\lambda} \tilde{b}_{\rho} )^{(1)} 
+ ( b_{\lambda}^{\dagger} b_{\eta}^{\dagger} )^{(1)} \cdot 
  ( \tilde{b}_{\eta} \tilde{b}_{\lambda} )^{(1)} \right. 
\nonumber\\
&& \hspace{1cm} \left.  
+ ( b_{\eta}^{\dagger} b_{\rho}^{\dagger} )^{(1)} \cdot 
  ( \tilde{b}_{\rho} \tilde{b}_{\eta} )^{(1)} \right]
\nonumber\\
&& + \sum_{L=0,2} d_{L} \,
( b_{\rho}^{\dagger} b_{\rho}^{\dagger} 
+ b_{\lambda}^{\dagger} b_{\lambda}^{\dagger} 
+ b_{\eta}^{\dagger} b_{\eta}^{\dagger} )^{(L)} \cdot ( {\rm h.c.} )^{(L)} ~, 
\label{HS4}
\ea
with $\tilde{b}_{k m}=(-1)^{1-m} b_{k -m}$ ($k=\rho$, $\lambda$, $\eta$) and $\tilde{s}=s$. 
The coefficients $\epsilon_{0}$, $\epsilon_{1}$, $u_{0}$, $u_{1}$, $v_{0}$, $a_{0}$, $a_{2}$, 
$c_{0}$, $c_{2}$, $c_{1}$, $d_{0}$ and $d_{2}$ parametrize the interactions. 
The Hamiltonian $H$ is diagonalized within the space of the totally symmetric representation 
$[N]$ of $U(10)$. 

Associated with the Hamiltonian, $H$, there are transition operators, $T$. Electromagnetic 
transition rates and form factors can all be calculated by considering the matrix elements 
of the operator
\ba
T &=& \mbox{e}^{-i q \beta D_{\eta,z}/X_D} ~,
\nonumber\\
D_{\eta,m} &=& (b_{\eta}^{\dagger} \times \tilde{s} - s^{\dagger} \times \tilde{b}_{\eta})^{(1)}_m ~,
\ea 
which is the algebraic image of the operator $\exp(iqr_{4,z})$ obtained from the full operator 
$\sum_{i=1}^4 \exp(i \vec{q} \cdot \vec{r}_i)$ by choosing the momentum transfer, $\vec{q}$, in 
the $z$-direction taken perpendicular to the base triangle in the direction of the 4th 
$\alpha$-particle and considering all particles to be identical (the coefficient $X_D$ is a 
normalization factor). 

The Hamiltonian of Eq.~(\ref{HS4}) with an appropriate choice of parameters can describe any 
dynamics of four-particles systems. In two cases, corresponding to the dynamic symmetries 
$U(10) \supset U(9)$ (harmonic oscillator) and $U(10) \supset SO(10)$ (deformed oscillator) 
the eigenvalues of the Hamiltonian $H$ of Eq.~(\ref{HS4}) can be obtained analytically. 
Here we discuss another situation, namely that of four particles at the vertices of a tetrahedron 
with ${\cal T}_d$ symmetry. The spectrum of a tetrahedral configuration can be obtained from the 
Hamiltonian of Eq.~(\ref{HS4}) by setting some coefficients equal to zero and taking specific 
linear combinations of others \cite{RB1,RB2}
\ba
H &=& \xi_{1} \, (R^{2} \,s^{\dagger} s^{\dagger} 
- b_{\rho}^{\dagger} \cdot b_{\rho}^{\dagger} 
- b_{\lambda }^{\dagger} \cdot b_{\lambda}^{\dagger} 
- b_{\eta}^{\dagger} \cdot b_{\eta}^{\dagger}) \, ( {\rm h.c.} )  
\nonumber\\
&& + \xi_2 \, \left[ ( -2\sqrt{2} \, b_{\rho}^{\dagger} \cdot b_{\eta}^{\dagger} 
+ 2 b_{\rho}^{\dagger} \cdot b_{\lambda}^{\dagger} ) \, ( {\rm h.c.} ) \right.
\nonumber\\
&& \hspace{1cm} + \left. ( -2\sqrt{2} \, b_{\lambda}^{\dagger} \cdot b_{\eta}^{\dagger} 
+ ( b_{\rho}^{\dagger} \cdot b_{\rho}^{\dagger} 
- b_{\lambda}^{\dagger} \cdot b_{\lambda}^{\dagger} )) \, ( {\rm h.c.} ) \right]
\nonumber\\
&& + \xi_3 \, \left[ ( 2 b_{\rho}^{\dagger} \cdot b_{\eta}^{\dagger}
+ 2\sqrt{2} \, b_{\rho}^{\dagger} \cdot b_{\lambda}^{\dagger} ) \, ( {\rm h.c.} ) \right.
\nonumber\\
&& \hspace{1cm} + ( 2 b_{\lambda}^{\dagger} \cdot b_{\eta}^{\dagger} 
+ \sqrt{2} \, ( b_{\rho}^{\dagger} \cdot b^{\dagger}_{\rho}  
- b_{\lambda}^{\dagger} \cdot b_{\lambda}^{\dagger} )) \, ( {\rm h.c.} )
\nonumber\\
&& \hspace{1cm} \left. + ( b_{\rho}^{\dagger} \cdot b_{\rho}^{\dagger} 
     + b_{\lambda}^{\dagger} \cdot b_{\lambda}^{\dagger} 
   - 2 b_{\eta}^{\dagger} \cdot b_{\eta}^{\dagger} ) \, ( {\rm h.c.} ) \right]
\nonumber\\
&& + \kappa_1 \, \vec{L} \cdot \vec{L} 
+ \kappa_2 \, (\vec{L} \cdot \vec{L} - \vec{I} \cdot \vec{I})^2 ~.  
\label{HTd}
\ea
Here $\vec{L}$ denotes the angular momentum in coordinate space ($x$, $y$, $z$) and $\vec{I}$ 
the angular momentum in the so-called `index' space $\rho$, $\lambda$, $\eta$. 

\begin{figure}
\centering
\includegraphics[width=3.2in]{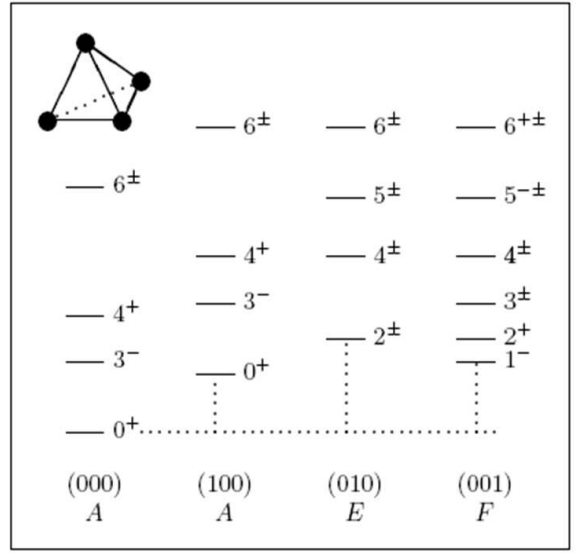}
\vspace{-0.5cm}
\caption{Schematic spectrum of a spherical top with tetrahedral symmetry 
and $\omega_1=\omega_2=\omega_3$. The rotational bands are labeled by 
$(v_1,v_2,v_3)$ (bottom). All states are symmetric under $S_4$.} 
\label{sphtop}
\end{figure}

The eigenvalues of $H$ of Eq.~(\ref{HTd}), given in terms of five parameters $\xi_1$, $\xi_2$, 
$\xi_3$, $\kappa_1$, $\kappa_2$ and the rigidity parameter $R^2$, cannot be obtained analytically. 
However, an approximate energy formula can be obtained by semiclassical methods 
($N \rightarrow \infty$ in $U(10)$). A tetrahedral 
configuration has three vibrational modes $v_1$, $v_2$ and $v_3$ labeled by their ${\cal T}_d$ 
symmetry. The vibration $v_1$ is the symmetric stretching (breathing mode) with $A$ symmetry. The 
vibration $v_2=v_{2a}+v_{2b}$ is the doubly degenerate vibration with $E$ symmetry and $a$, $b$ 
components. The vibration $v_3=v_{3a}+v_{3b}+v_{3c}$ is the triply degenerate vibration with $F$ 
symmetry and $a$, $b$, $c$ components. Since the tetrahedral group ${\cal T}_d$ is isomorphic to 
the permutation group $S_4$, the vibrations can also be labeled by representations of $S_4$: 
$[4] \sim A$, $[22] \sim E$, $[31] \sim F$. The vibrational spectrum is 
\ba
E_{\rm vib} = \omega_{1}(v_{1}+\frac{1}{2}) 
+ \omega_{2}(v_{2}+1) + \omega_{3}(v_{3}+\frac{3}{2}) ~, 
\label{Evib}
\ea
with frequencies 
\ba
\omega_{1} = 4NR^{2} \xi_{1} ~, \;  
\omega_{2} = \frac{8NR^{2}}{1+R^{2}} \xi_{2} ~, \; 
\omega_{3} = \frac{8NR^{2}}{1+R^{2}} \xi_{3} ~.
\ea
For rigid configurations, $R^2=1$ and $\omega_i=4N \xi_i$ (with $i=1,2,3$). 
The rotational states built on top of each vibration have angular momenta and 
parities determined by the invariance of the Hamiltonian under $S_4$, {\it i.e.} 
all states must be symmetric under $S_4$. 
As a consequence, states with $A$ symmetry have angular momentum and parity 
$L^P=0^+$, $3^-$, $4^+$, $6^{\pm}$, $\ldots$, while states with $E$ symmetry have 
$L^P=2^{\pm}$, $4^{\pm}$, $5^{\pm}$, $6^{\pm}$, $\ldots$, and states with $F$ symmetry 
have $L^P=1^-$, $2^+$, $3^{\pm}$, $4^{\pm}$, $5^{-\pm}$, $6^{+\pm}$, $\ldots$.  
Note the unusual composition of the rotational band built on the ground state 
($A$ symmetry). This angular momentum content is in agreement with that observed 
in molecules with ${\cal T}_d$ symmetry (see page 450 of \cite{Herzberg}). This content 
has also been derived in \cite{Kameny} for applications to nuclei. The rotational spectrum 
depends on $L$ and $I$, and on the parameters $\kappa_1$ and $\kappa_2$ in 
Eq.~(\ref{HTd}). However, for the excitations of a rigid spherical top, $L=I$, 
the last term in Eq.~(\ref{HTd}) does not contribute, and the rotational energies 
are given by $E_{\rm rot} = \kappa_{1} \, L(L+1)$. 
The combined rotation-vibration spectrum of a tetrahedral cluster is shown in 
Fig.~\ref{sphtop}. 

\begin{figure}
\centering
\includegraphics[width=3.2in]{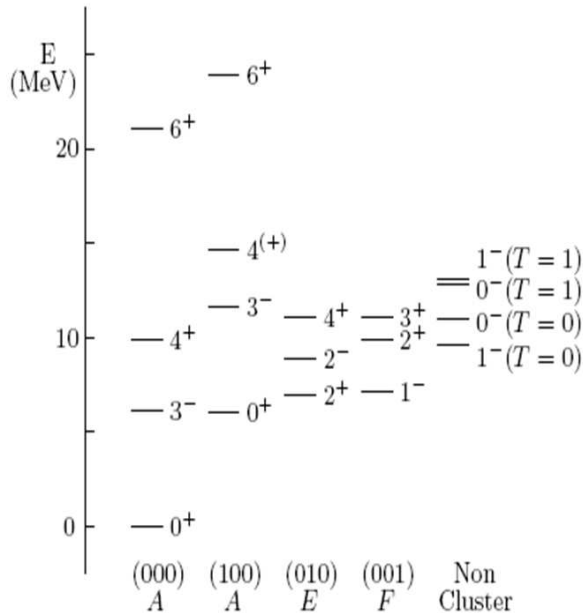}
\vspace{-0.5cm}
\caption[]{The observed spectrum of $^{16}$O \cite{NDS}. The levels are organized in columns 
corresponding to the ground state band and the three vibrational bands with $A$, $E$ and $F$ 
symmetry of a spherical top with tetrahedral symmetry. The last column shows 
the lowest non-cluster levels.} 
\label{O16}
\end{figure}

The matrix elements of the electromagnetic transition operator, $T$, and form factors 
in the ACM are the representation (Wigner) matrix elements of $U(10)$. We have derived 
closed forms of these in the $U(9)$ and $SO(10)$ dynamic symmetries and in the large $N$ 
limit for the spherical top with tetrahedral symmetry. This constitutes an important 
new result of the ACM. In the spherical top case discussed here, the form factors for 
transitions along the ground state band $(0,0,0)A$ are given by spherical Bessel 
functions, $F_L(0^+ \rightarrow L^P;q)=c_L j_L(q \beta)$. The coefficients $c_L$ for 
the first few states are $c_0^2=1$, $c_3^2=35/9$, $c_4^2=7/3$ and $c_6^2=416/81$ 
for the $L^P=0^+$, $3^-$, $4^+$ and $6^+$, respectively. 
The transition probabilities $B(EL)$ can be extracted from the form factors  
in the long wavelength limit 
\ba
B(EL;0 \rightarrow L) = \left(\frac{Ze\beta^{L} }{4}\right)^{2} 
\frac{2L+1}{4\pi} \left[ 4+12P_{L}(-\frac{1}{3}) \right] 
\label{BEL}
\ea
The form factors and $B(EL)$ values only depend on the parameter $\beta$, the distance 
of each $\alpha$ particle from the center of the tetrahedral configuration., and on the 
${\cal T}_d$ symmetry which gives the coefficients $c_L$. By extracting the value of $\beta$ 
from the elastic form factor measured in electron scattering, one can thus make a model 
independent test of the symmetry. 

Whereas $L$ is an exact symmetry of $H$, $I$ is not. If $L \neq I$, perturbations must 
be added. The algebraic model allows one to study these perturbations quantitatively by 
diagonalizing the Hamiltonian $H$ of Eq.~(\ref{HTd}) in an appropriate basis. A convenient 
basis to construct states with good permutation symmetry $S_4$ is the 9-dimensional 
harmonic oscillator basis \cite{KM} corresponding to the reduction 
$U(10) \supset U(9) \supset U(3) \otimes U(3) \otimes U(3)$. 
We have constructed a set of computer programs to calculate energies and electromagnetic 
transition rates in this basis. 

\begin{figure}
\centering
\includegraphics[width=3in]{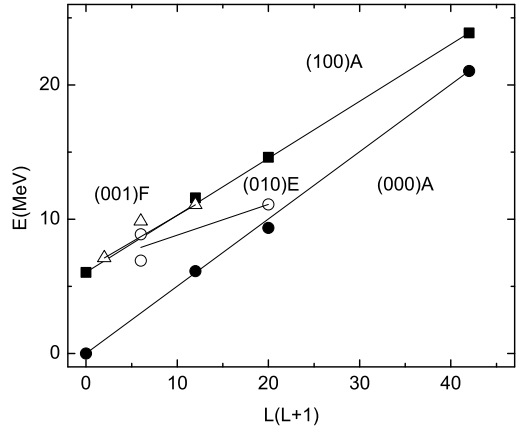}
\vspace{-0.5cm}
\caption[]{The excitation energies of cluster states in $^{16}$O 
plotted as a function of $L(L+1)$: closed circles for the ground state band, 
closed squares for the $A$ vibration, open circles for the $E$ vibration 
and open triangles for the $F$ vibration.} 
\label{bands}
\end{figure}

Our derivation of the spectrum of clusters with ${\cal T}_d$ symmetry can be used to study 
cluster states in $^{16}$O. The observed experimental spectrum of $^{16}$O is shown in 
Fig.~\ref{O16}. It appears that a rotational ground state band with angular momenta 
$L^P=0^+$, $3^-$, $4^+$, $6^+$ has been observed with moment of inertia such that 
$\kappa_1=0.511$ MeV. It appears also that all three vibrations, $A$, $E$ and $F$, have 
been observed with comparable energies, $\sim 6$ MeV, as one would expect from Eq.~(\ref{Evib}) 
if $\xi_1=\xi_2=\xi_3$. A rotational band with $0^+$, $3^-$, $4^+$, $6^+$ appears also to have 
been observed for the $A$ vibration $(1,0,0)$ (breathing mode). This band is similar in nature 
to the band built on the Hoyle state in $^{12}$C and recently observed \cite{Itoh,Freer,Gai}. 
It has a moment of inertia such that $E=0.463 \; L(L+1)$ MeV. The moment of inertia of the $A$ 
vibration is larger than that of the ground state due to its nature (breathing vibration). 
The situation is summarized in Fig.~\ref{bands}. 
The observed spectrum has perturbations. The most notable perturbation is the splitting of the 
$2^{\pm}$ states of the $E$ vibration. This cannot be simply described by the formula 
$E \propto L(L+1)$ and requires a diagonalization of the full Hamiltonian. 

Having identified the cluster states, one can then test the ${\cal T}_d$ symmetry by means 
of the electromagnetic form factors and $B(EL)$ values. We extracted the value of $\beta$ 
from the first minimum in the elastic form factor \cite{Sick}, obtaining $\beta=2.0$ fm. 
Table~\ref{BELtable} 
shows the results for the $B(EL)$ values. The ${\cal T}_d$ symmetry appears to be unbroken in the 
ground state band of $^{16}$O. We also investigated the electromagnetic decays of the vibrational 
bands $(1,0,0)A$, $(0,1,0)E$ and $(1,0,0)F$. For these bands the ${\cal T}_d$ symmetry appears 
to be broken and, in addition, they decay mostly by $E2$ quadrupole transitions. For $E2$ transitions 
the simple analytic formula of Eq.~(\ref{BEL}) does not apply, since the $E2$ operator is not in the 
representation $A$ of ${\cal T}_d$ as the $E3$, $E4$ and $E6$ operators and hence can connect 
different representations. A full account of these transitions will be given in a forthcoming 
longer publication \cite{RB3}. 

\begin{table}
\caption{Comparison of theoretical and experimental $B(EL)$ values in e$^2$fm$^{2L}$ 
and $E_{\gamma}$ values in keV, along the ground state band. The theoretical $B(EL)$ values 
are obtained from Eq.~(\ref{BEL}), and the $E_{\gamma}$ values are obtained from 
$E=0.511 \, L(L+1)$ MeV. The experimental values are taken from \cite{NDS}.}
\label{BELtable}
\begin{tabular}{cccccc}
\hline
\hline
\noalign{\smallskip}
$B(EL;L^P \rightarrow 0^+)$ & Th & Exp & $E_{\gamma}(L^P)$ & Th & Exp \\
\noalign{\smallskip}
\hline
\noalign{\smallskip}
$B(E3;3_1^- \rightarrow 0_1^+)$ &  181 & $205 \pm  10$ & $E_{\gamma}(3_1^-)$ &  6132 &  6130 \\
$B(E4;4_1^+ \rightarrow 0_1^+)$ &  338 & $378 \pm 133$ & $E_{\gamma}(4_1^+)$ & 10220 & 10356 \\
$B(E6;6_1^+ \rightarrow 0_1^+)$ & 8245 &               & $E_{\gamma}(6_1^+)$ & 21462 & 21052 \\ 
\noalign{\smallskip}
\hline
\hline
\end{tabular}
\end{table}

Cluster states represent only a portion of the full spectrum of states. They are obtained by 
assuming that the $\alpha$ particles have no internal excitation. 
At energies of the order of the shell gap, $\sim 16$ MeV in $^{16}$O, one expects to have 
non-cluster states, and thus the spectrum to be composed of cluster states immersed into a 
bath of non-cluster states. Assigning states to cluster or non-cluster above this energy 
is a difficult task. We note, however, that the tetrahedral structure in Fig.~\ref{sphtop} 
has {\em no} $0^-$ state and only one $1^-$ state in the $F$-vibration. Thus $0^-$ states 
are clearly non-cluster states. Also with $\alpha$ particles one cannot form $T=1$ states. 
These states are obviously non-cluster. In Fig.~\ref{O16}, we have assigned the states 
$L^P=1^-$, $0^-$ ($T=0$) at $E=9.585$ MeV and $10.957$ MeV and 
$L^P=0^-$, $1^-$ ($T=1$) at $E=12.796$ MeV and $13.090$ MeV, as the shell model 
configuration $1p_{1/2}^{-1} 2s_{1/2}$. The shell model states $1p_{1/2}^{-1} 1d_{5/2}$ 
with $L^P=2^-$, $3^-$ and $T=0$ and $T=1$ can also be easily identified but they are 
not shown in Fig.~\ref{O16} not to overcrowd the figure. For the same reason, we do not 
show in Fig.~\ref{O16} other states with $L^P=4^{\pm}$, $5^-$, $6^{\pm}$, $\ldots$ which 
can be assigned to cluster configurations. 

An important question is the shell-model description of cluster states. It was suggested long 
ago \cite{Feshbach,Brown} that the state at $6.049$ MeV is a $4p-4h$, while the state at 
$7.116$ MeV is a $5p-5h$. In view of the recent developments of large-scale shell model calculations 
and of the no-core shell model it would be interesting to study once more the shell model 
description of the states in Fig.~\ref{O16}.

Very recently, also an {\it ab initio} lattice calculation of the spectrum and structure of 
$^{16}$O has been reported \cite{lattice}. This calculation confirms the tetrahedral structure 
of the ground state of $^{16}$O in agreement with our findings. For the excited states, $0_2^+$ 
and $2_1^+$ instead, a square configuration is suggested. This would imply a large breaking 
of the ${\cal T}_d$ symmetry for the vibrations in Fig.~\ref{O16}. Although we expect the 
${\cal T}_d$ symmetry to be broken for the vibrational states due to the near degeneracy of them, 
$\xi_1=\xi_2=\xi_3$, {\it i.e.} even a small breaking term in $H$ may cause a large mixing, we 
nonetheless feel at this stage that our interpretation of the excited states of $^{16}$O as 
vibrations provides a good starting point for further studies. Algebraic methods are quite 
general, and as shown in Ref.~\cite{Larese}, they can accommodate all sorts of configurations 
of four particles, including configurations with ${\cal T}_d$, ${\cal D}_{3h}$ and 
${\cal D}_{4h}$ (square) symmetry. 
In connection with tetrahedral configurations in nuclei, we mention here also the work of 
\cite{Rae} in light nuclei and \cite{Dudek} in heavy nuclei for which, however, there is no 
experimental confirmation.  

In conclusion, we have introduced an algebraic model capable of describing the full dynamics  
of four-body clusters. Within this model we have rederived the spectrum of a spherical top
with tetrahedral symmetry, and confirmed the evidence for the occurrence of this symmetry 
in the low-lying spectrum of $^{16}$O presented long ago by Kameny \cite{Kameny} and 
Robson \cite{Robson}. An analysis of the $B(EL)$ values along the ground state band provide 
an even stronger evidence for ${\cal T}_d$ symmetry than the energies. 
Another crucial aspect is the development of the $U(10)$ ACM for four-body clusters which 
allows a detailed description of energies, electromagnetic transition rates, form factors and 
$B(EL)$ values. We hope that the results in this paper will stimulate further experimental work 
on the structure of $^{16}$O. Finally, the results presented here in conjunction with 
those in $^{12}$C emphasize the occurrence of $\alpha$-cluster states in light nuclei. 

This work was supported in part by research projects from DGAPA-UNAM and CONACyT, Mexico
and in part by U.S. D.O.E. Grant DE-FG02-91ER40608.

\end{document}